# Understanding discrepancies in the coverage of OpenAlex: the case of China

Mengxue Zheng[1], Lili Miao[2], Yi Bu[3,4], Vincent Larivière[5,6]

[1]School of Information Management, Wuhan University, Wuhan 430072, China.

[2]Center for Complex Networks and Systems Research, Luddy School of Informatics, Computing, and Engineering, Indiana University Bloomington, IN 47408, USA.

[3]Department of Information Management, Peking University, Beijing 100871, China.

[4]Center for Digital Intelligence Science and Education Research, Peking University Chongqing Research Institute of Big Data, Chongqing 401332, China.

[5]École de bibliothéconomie et des sciences de l'information, Université de Montréal, Montréal, QC. H3C 3J7, Canada.

[6]Observatoire des Sciences et des Technologies (OST), Centre Interuniversitaire de Recherche sur la Science et la Technologie (CIRST), Université du Québec à Montréal, Montréal, QC. H3C 3P8, Canada.

**Abstract**

Citations indexes play a crucial role for understanding how science is produced, disseminated, and used. However, these databases often face a critical trade-off: those offering extensive and high-quality coverage are typically proprietary, whereas publicly accessible datasets frequently exhibit fragmented coverage and inconsistent data quality. OpenAlex was developed to address this challenge, providing a freely available database with broad open coverage, with a particular emphasis on non-English speaking countries. Yet, few studies have assessed the quality of the OpenAlex dataset. This paper assesses the coverage, by OpenAlex, of China's papers, which shows an abnormal trend, and compares it with other countries that do not have English as their main language. Our analysis reveals that while OpenAlex increases the coverage of China's publications, primarily those disseminated by a national database, this coverage is incomplete and discountinuous when compared to other countries' records in the database. We observe similar issues in other non-English-speaking countries, with coverage varying across regions. These findings indicate that, although OpenAlex expands coverage of research outputs, continuity issues persist and disproportionately affect certain countries. We emphasize the need for researchers to use OpenAlex data cautiously, being mindful of its potential limitations in cross-national analyses.

## 1 Introduction

Bibliographic databases containing the metadata of research publications are crucial for advancing our understanding of science and innovation. For past decades, the research community has primarily relied on two leading sources—Web of Science (WoS) and Scopus—due to their extensive coverage and continuous data improvements and updates (Zhu & Liu, 2020). However, both datasets exhibit inherent biases. Developed and maintained by Western, Anglophone for profit entities, they tend to overrepresent research from Western countries and English-language publications and underrepresent research output from other regions and non-English languages (Alperin et al., 2024; Archambault et al., 2006; Mongeon & Paul-Hus, 2016; Priem et al., 2022; Visser et al., 2021). Additionally, they prioritize established journals, often excluding alternative scholarly outputs such as preprints, conference proceedings, and papers published in repositories (Mongeon & Paul-Hus, 2016). Furthermore, WoS and Scopus operate on proprietary platforms that require costly subscriptions, limiting access for financially constrained researchers and institutions (Boudry et al., 2019; Heeks, 2022). The proprietary nature of these datasets also restricts data sharing, as researchers are bould by licenses that prohibit free data exchange and transparency in collaborative projects.

In response to these limitations, OpenAlex was developed by the nonprofit organization OurResearch to provide an inclusive, publicly accessible bibliographic dataset for the global research community (Alperin et al., 2024; Haupka et al., 2024; Priem et al., 2022). OpenAlex provides a freely accessible API as well as monthly updated data snapshots, making it available to researchers at no cost. Compared to WoS and Scopus, OpenAlex provides more comprehensive coverage, not only in journal articles but also in diverse outputs such as conference proceedings, preprints, and institutional repositories. Additionally, OpenAlex seeks to address the English-dominant bias present in other major datasets by including more non-English works (Céspedes et al., 2024), particularly from the Global South, thereby enhancing the visibility of diverse research contributions (Alperin et al., 2024). These characteristics make OpenAlex a valuable resource for a broad range of analyses, from exploring connections between science and innovation to conducting more balanced evaluations of regional and national scientific development (Akbaritabar et al., 2024; Haunschild & Bornmann, 2024; Massri et al., 2023; Woelfle et al., 2023).

However, OpenAlex is not without errors. In addition to data quality issues (Alperin et al., 2024) such as missing affiliation or citation information, a distinctive trend appears in the data concerning publications from China. Rather than following a smooth, exponential growth pattern, the number of China's publications indexed in OpenAlex shows a sharp rise between 2000 and 2010, reaching parity with the United States by 2011, followed by a steep decline between 2012 and 2016 (Alperin et al., 2024). Given the critical importance of accurately capturing national scientific output for understanding scientific systems, it remains unclear whether this trend in OpenAlex reveals unique insights absent from other datasets or is simply an artifact based in indexing inconsistency. Motivated by these observations, this study investigates the underlying reasons for the observed fluctuations in China's publication data in OpenAlex and explores the implications of using OpenAlex for research on national scientific output.

## 2 Related Works

OpenAlex inherited most of its data from the Microsoft Academic Graph (MAG), it has maintained bibliographic information on works while improving the categorization of document types, appearing better suited for bibliometric research than MAG (Scheidsteger & Haunschild, 2023). As widely regarded as one of the most promising inclusive open science databases(Alperin et al., 2024; Céspedes et al., 2025), OpenAlex has gained increasing attention in studies concerning its coverage and data quality.

Several studies have compared the coverage of OpenAlex with established, inclusive databases—WoS and Scopus. A notable study by Alperin et al. (2024) compares the coverage and metadata of publications from 2000 to 2022 between OpenAlex (using the November 21, 2023 snapshot) and Scopus. Their findings highlight that OpenAlex provides broader coverage of publications, but also has some limitations: many works lack identifiable country data, a significant number lack clear source type information, and it does not include a complete citation graph for the indexed works. Furthermore, no published work to date has examined the quality of language detection in OpenAlex. Culbert et al. (2024) examined a shared corpus of 16,788,282 publications (from 2015 to 2022) indexed by OpenAlex, WoS, and Scopus. They found that OpenAlex achieves comparable or slightly higher reference coverage than WoS and Scopus. Thelwall & Jiang (2025) also confirmed OpenAlex's suitability for citation analysis when compared with Scopus. Additionally, OpenAlex is found to index a wider range of open access journals compared to Scopus and WoS, with a more balanced disciplinary representation, especially in the social sciences and humanities (Maddi et al., 2024). It also provides the most extensive coverage of African publications when compared to WoS and Scopus (Alonso-Alvarez & Eck, 2025). However, OpenAlex still faces challenges with less precise classifications of document types compared to Scopus and WoS (Haupka et al., 2024).

In terms of data quality, Céspedes et al. (2025) investigate the accuracy of language metadata in OpenAlex, focusing on 11 key languages out of 55. Their study compares OpenAlex with WoS and uses a stratified sample of 6,836 articles published between 2000 and 2020. The results indicate that 98% of works indexed in OpenAlex contain language information, but approximately 14.7% are inaccurately labeled. Similarly, Zhang et al. (2024), based on the October 10, 2022 snapshot of OpenAlex, found that, while the proportion of articles with full institutional information has generally increased, over 60% of OpenAlex articles still lack institutional data, particularly in the social sciences and humanities. The underrepresentation of China's institutions is particularly notable due to missing institutional data. Despite these challenges, OpenAlex performs well in terms of African publication and author information, though it lags behind Scopus and WoS in institutional, reference, and funding information coverage(Alonso-Alvarez & Eck, 2025).

In response to these findings, we proposed hypotheses for the reasons behind the puzzling trend of China's publications that we encountered when analyzing OpenAlex data, and provided related results in this paper.

### 3 Data and Methods
The OpenAlex snapshot used in this study was released on June 30th, 2024. The dataset contains a total of 256,997,006 scholarly works covering the period from 1400 to 2024. OpenAlex encompasses a wide range of research outputs typically excluded from conventional bibliographic datasets, such as datasets, peer-review reports, and books. Since the focus of our study is to evaluate OpenAlex's coverage of research output, we restrict our analysis to articles, preprints, and conference proceedings. Additionally, we narrow the analysis to the top 10 most productive countries, as ranked by the Nature Index, and the time frame to 2000-2023 to focus on a more recent coverage. Country information was derived from the "authorships"-"institutions"-"country_code" field, which is based on the institutional addresses listed by the authors. For this analysis, the country associated with each institution of each author is counted. For a paper with multiple countries listed, we apply the full counting method, assigning the paper to each participating country. The language of each publication is specified in the 'language' field. For this analysis, only publications labeled as written in English in OpenAlex are included. The criteria used in manual check of this study for determining the actual language of publication are based on full text. The indexed source of each publication is identified using the "primary_location"-"landing_page_url" field. After removing duplicate entries, our final dataset comprises 42,957,650 publications authored by the top 10 countries between 2000 and 2023. Given that previous research has shown issues with OpenAlex in correctly identifying the institutions of papers (Zhang et al., 2024), and considering that a major advantage of the OpenAlex dataset is its broader coverage compared to other prevalent bibliographic databases, thanks to its more inclusive data collection mechanism, we propose two hypotheses for the mechanism underlying the abnormal trend in the data regarding publications from China:

- Hypothesis 1. The abnormal trend in China's publications in OpenAlex is due to incorrect identification of the country of affiliated institutions.
- Hypothesis 2. The abnormal trend in China's publications in OpenAlex is due to changes in source coverage over time.

## 4 Results
### 4.1 Initial checks
We first assess OpenAlex's coverage for the top 10 most productive countries (Figure 1a). The overall trend from OpenAlex aligns with existing studies, showing that the United States and China are the largest producers of publications, while the other eitght other countries have much lower outputs. However, a notable pattern that emerges from the OpenAlex data is China's publication output, which experiences dramatic fluctuations, echoing Alperin et al. (2024)'s results. From 2000 to 2011, the number of publications from China grow rapidly, reaching levels similar to the United States by 2011. However, the trend then sharply declines, reaching a low point in 2016. This level of fluctuation is not observed in other top-10 countries.

To determine whether the fluctuation in China's publication output is an artifact of the OpenAlex data or reflects its actual performance, we compare the number of China's publications indexed in WoS and Scopus. For Web of Science and Scopus, country information was derived from the institutional addresses listed by all contributing authors. Language information was taken from the 'language' field provided by each database, which relies on metadata supplied by publishers. For a more comprehensive comparison, we also include the publication performance of the United States. As shown in Figure 1b, both China and the United States display a steady growth trend in WoS and Scopus, without the fluctuations observed in OpenAlex. These fluctuations end in 2016, and from that year onwards, the English-language China's output in OpenAlex is almost identical to that observed in WoS and Scopus.

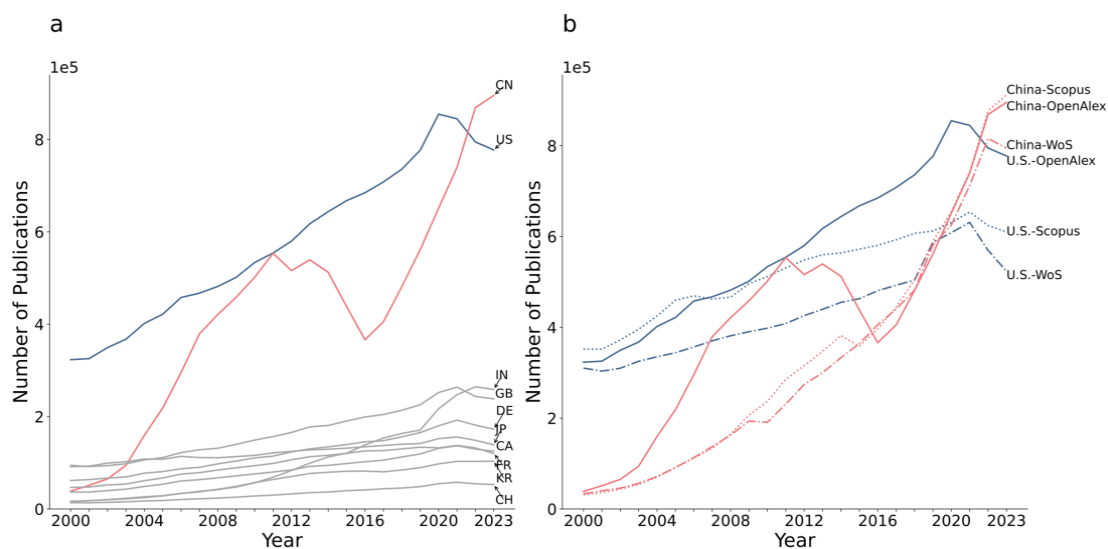

**Figure 1.** Number of publications from China indexed in OpenAlex shows a sharp, fluctuating pattern. (a) The number of publications for the top 10 countries indexed by OpenAlex. (b) Number of publications of China and the United States indexed by OpenAlex, WoS, and Scopus.

**4.2 Publication type**

To identify which type of publications contribute to the fluctuation, we further break down the data into publication types—namely, articles, preprints, and conference proceedings—and measure their corresponding growth trends. The results show that, while the majority of works are research articles, the fluctuation stems from the indexed journal articles. Therefore, in the subsequent analysis, we focus on the coverage quality of journal articles.

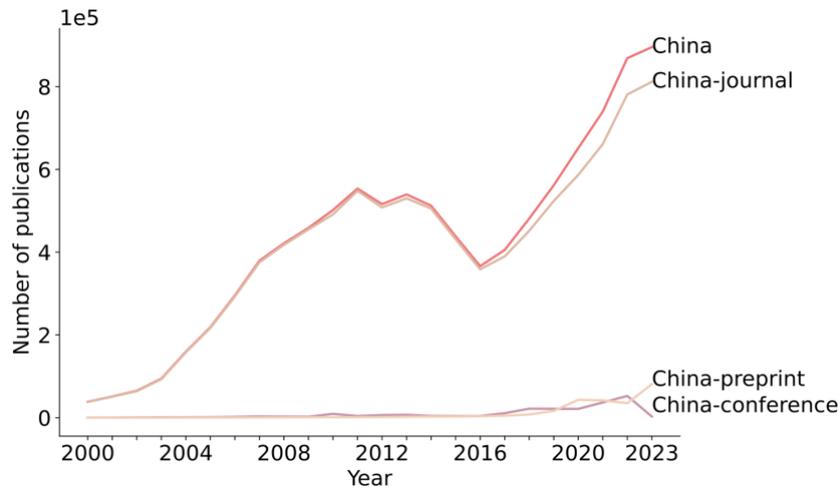

**Figure 2.** Number of publications from China indexed by OpenAlex over years by publication type.

To uncover the mechanism behind the fluctuation regarding the number of journal publications, we first examine Hypothesis 1. This hypothesis indicates that the bump between 2000 and 2016 of China is attributable to incorrect country identification of affiliated institutions, leading to publications from other countries being misidentified as China's publications. This explanation is plausible given that it is unlikely for the number of China's English-language publications in 2011 to have genuinely matched that of the United States. To test this hypothesis, we analyze institutions with publications in 2011 that are absent in 2016, as the fluctuation peaked in 2011 and declined sharply by 2016. The absence of these institutions in 2016 may indicate that OpenAlex had corrected their affiliated country information by that time. In total, we find that 301 China's institutions had indexed papers in 2011 but were missing in 2016. We manually cross-checked the actual country information for the top 50 institutions with the highest number of indexed publications from this group. All of these institutions are confirmed to be located in China, suggesting that the bump is not attributable to country misclassification of non-Chinese institutions as Chinese ones.

We then test Hypothesis 2 to understand whether the observed fluctuation is related to changes in journal coverage. To investigate this, we analyze the number of journals indexed in 2011 that were no longer indexed in 2016. We find that 4,506 journals were indexed in 2011 but not in 2016, accounting for approximately 40.56% of the total publications in 2011. Upon further examining the primary locations of the paper from the missing journals, we find that a significant portion (204,442 publications, 92%) were indexed from the China National Knowledge Infrastructure (CNKI)—the largest bibliographic database in China. Next, we focus on the top 50 journals (out of 7.903 journals indexed in both 2011 and 2016) that experienced the highest reduction rate in publications. These journals had 55,536 papers indexed in 2011, accounting for 10.13% of total publications that year, but only 114 papers in 2016. Notably, of these 50 journals, 49 had their articles indexed by OpenAlex from CNKI. By 2016, the total number of CNKI-sourced papers dropped to 5,783 (1.6%). These results suggest that papers indexed from CNKI may contribute to the fluctuation in China's publications in OpenAlex.

**4.3 CNKI and OpenAlex**

To further examine whether the heterogeneous coverage of publications from CNKI is the main factor behind the fluctuation in China's indexed papers, we measure the number of papers indexed from CNKI over the years. The results show that OpenAlex has a very unstable coverage of papers from CNKI. From 2003 to 2011, a steady proportion of China's publications (around 60%-70%) were collected from CNKI. However, this proportion dropped to 34.7% in 2014, and by 2016, the proportion of papers from CNKI had decreased to nearly zero (see Figure 3a). Figure 3b shows that excluding China's publications indexed from CNKI eliminates the fluctuations in China's publication trends.

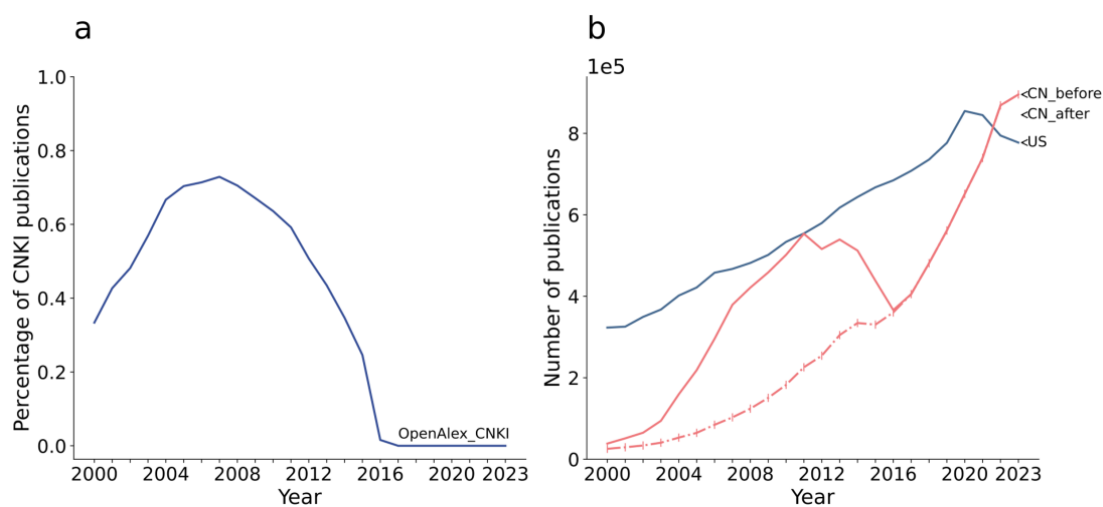

**Figure 3**. Fluctuations in China's publication numbers disappear when CNKI-sourced publications are removed. (a) Proportion of CNKI-sourced publications within China's publications indexed in OpenAlex. (b) Number of China's publications indexed by OpenAlex per year: before/after removing publications of China collected from CNKI, compared with the United States.

Although OpenAlex has shown unstable coverage of papers from CNKI over the years (see Figure 3a), it remains valuable to include publications from CNKI, as it could serve as a complementary data source to compensate for the lack of national data in international databases. However, given that CNKI is a comprehensive national database that primarily provides full-text publications written in Chinese, we did not expect the coverage of CNKI to affect the number of English-language publications in OpenAlex substantially. We randomly sample 100 papers indexed from CNKI in years 2000, 2003, 2006, 2009, 2011, 2013, 2016; these papers are labeled as written in English by OpenAlex; a manual check reveals that 99% of these papers are written in Chinese. A chi-square test reveals a significant difference ($p < 0.05$) in the full-text language distribution between CNKI and non-CNKI publications (Appendix A, Table A1). To further assess CNKI's coverage in OpenAlex before 2016, when OpenAlex still collected data from CNKI, we compare the number of publications indexed by CNKI with those collected by OpenAlex from CNKI, especially for two leading universities in China— Tsinghua University and Peking University from 2000 to 2015. Publications indexed by CNKI are limited to Chinese journal articles with available full texts, as we find the majority of works that OpenAlex indexed from CNKI are such type. The results show that, for both universities, the proportion of CNKI articles captured by OpenAlex remains relatively low and inconsistent across the years.

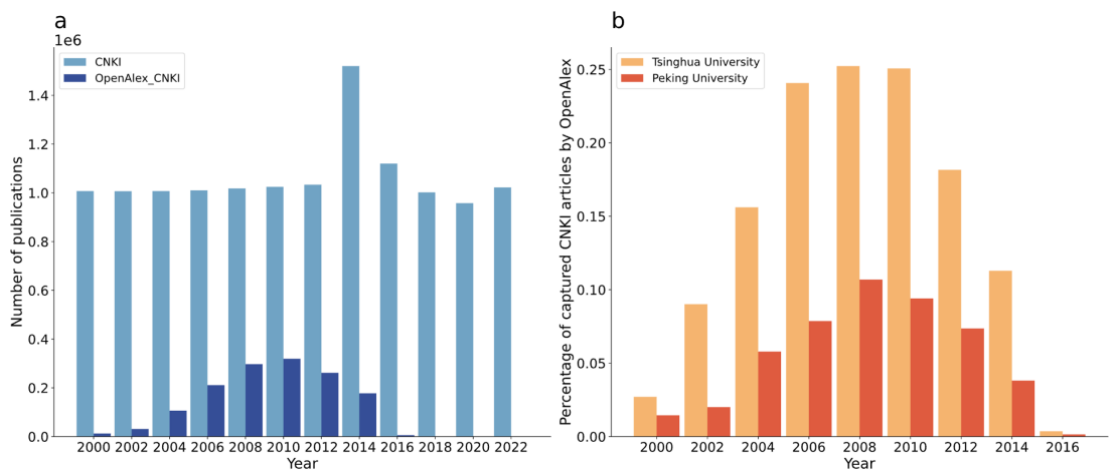

**Figure 4**. Coverage of CNKI articles in OpenAlex over years. (a) Number of China's articles in CNKI v.s. Number of China's publications indexed from CNKI in OpenAlex. (b) Proportion of CNKI articles captured by OpenAlex for Tsinghua University and Peking University.

Then we shift our focus to Chinese-language works indexed in OpenAlex for a while, as it claims to have better coverage of non-English publications (Céspedes et al., 2024), to explore whether improvements in language detection might explain the observed decline in English-language publications from China. The results show that 5,140 publications are labeled as Chinese language between 2000 and 2023, with 46 of these collected from CNKI, which suggests that OpenAlex has largely ceased indexing CNKI publications. Therefore, while including publications from national databases like CNKI could provide valuable insights, the low representativeness and unstable coverage of these publications in OpenAlex make them unsuitable for reliable analysis.

**4.4 Incorrect language identification and incomplete affiliation information**
Examination of the consistency of metadata for papers from CNKI indexed in OpenAlex reveal two additional issues: incorrect language identification and incomplete affiliation information. Although the language of a work is identified based on the abstract and title rather than the full text, the results show a significant discrepancy between the language identified from the abstract and title and the actual language of the full text. Before 2016, a large proportion of publicationss written in Chinese were incorrectly labeled as English by OpenAlex (see Figure 5). Almost all of these papers were sourced from CNKI. While national databases may provide English titles and abstracts, relying on this information to infer the language of the full text leads to misidentification, especially for works from national databases. Additionally, out of 1,621 papers collected from CNKI, OpenAlex failed to correctly extract the full list of authors. In these cases, only the first author was captured, with the remaining authors entirely missing from the OpenAlex metadata.

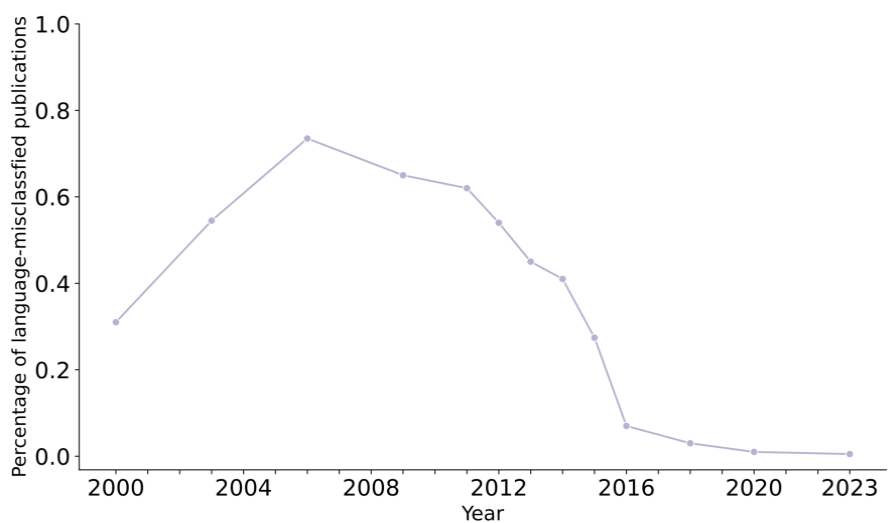

**Figure 5.** Proportion of non-English publications from China that are misclassified as English language by OpenAlex. This figure presents the results of a manual check of a random sample of 200 publications for each selected year; these publications are labeled as written in English, typed as articles, preprints, or conference proceedings, with author/s affliated with China's institutions in OpenAlex. The years selected for the analysis are 2000, 2003, 2006, 2009, 2011, 2012, 2013, 2014, 2015, 2016, 2018, 2020, and 2023, providing a comprehensive overview of the language misclassification trends across time.

**4.5 Analysis of other countries beyond China**

Given that including publications written in non-English languages provides critical complementary information for current bibliographic datasets, we expand our analysis from China's national coverage to other countries whose official language is also not English. We select South Korea and France due to their significant scientific output, representing non-English speaking countries in Asia and Europe, respectively. Additionally, we notice that Indonesia ranks among the top 10 most productive countries in OpenAlex, based on the number of publications in 2022, which contrasts sharply with the top 10 productive countries in other datasets (National Science Board, 2024). Indonesia is also known for its strong open access culture(Simard et al., 2024). As a result, we include Indonesia in the sample. To investigate national coverage, we randomly sample 200 works from each country in years 2000, 2006, 2012, 2018, 2023;

these works are labeled as written in English, typed as articles, preprints, or conference proceedings, with author/s affliated with institutions in the respective country, according to OpenAlex. In a few cases, full texts are inaccessible. For the accessible works, we manually examined the proportion of language-misclassified works where the full text is not in English.

The results reveal that although France and South Korea have their own languages and national databases, such as Hyper Articles en Ligne of France, Korean Studies Information Service System and KoreaScience of South Korea, and some publications in those databases are indexed in OpenAlex, the proportion of papers indexed in OpenAlex that are not written in English is low. In contrast, a relatively high proportion of papers produced by researchers from Indonesia are not written in English. Among these Indonesian papers, 632 (63.2%) are identified as open access. These results demonstrate that metadata issues in OpenAlex are not isolated to China.

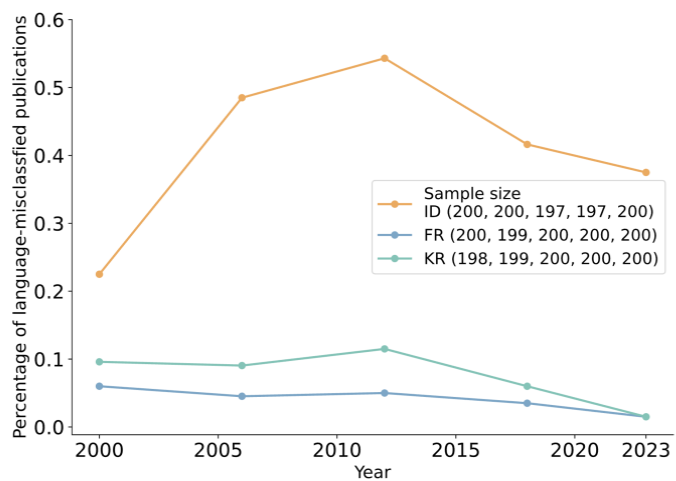

**Figure 6**. Proportion of non-English publications of Indonesia, France, or South Korea that misclassified as English language by OpenAlex. This figure presents the results of a manual check of a random sample of 200 publications of per country for each selected year.

We then compare the number of publications indexed by OpenAlex and WoS, a very established and commonly-adopted bibliographic database, for the top 10 countries, as well as Indonesia and Russia for their complex publishing environments or linguistic situations. The results reveal inconsistencies in national coverage between the two databases, particularly for China, Indonesia, and Russia. As shown in Figure 7, China is the only country for which OpenAlex indexed more publications than WoS in the early years, but fewer publications in the late years (from 2016 to 2020), while many countries have unique characteristics.

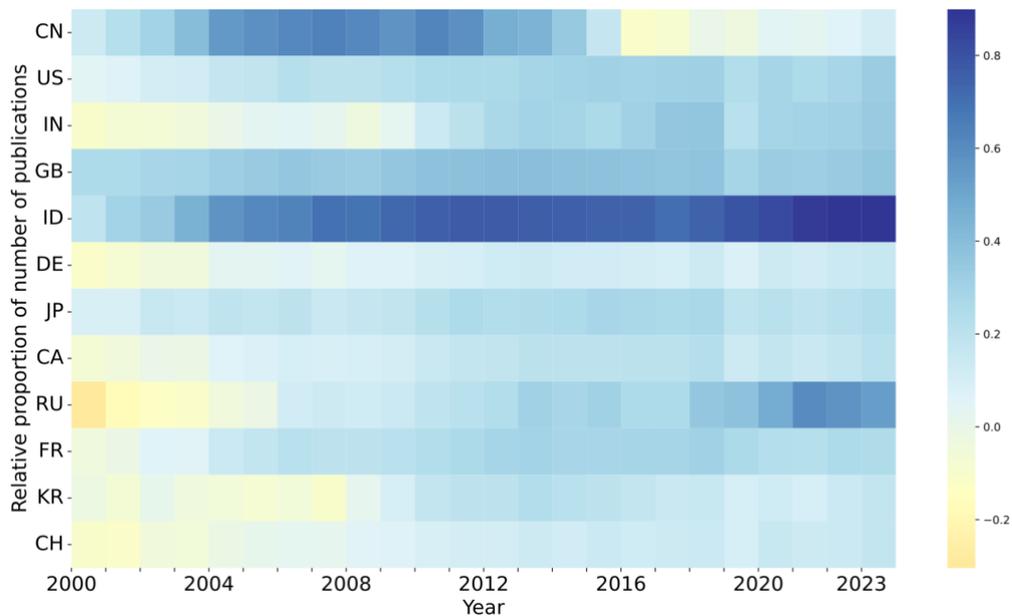

**Figure 7**. Relative proportions of publications indexed by OpenAlex and WoS for 12 countries. Positive values indicate more publications in OpenAlex than WoS, and *vice versa*.

## 5 Discussion

The accuracy of bibliographic databases is a key issue for further quantitative studies. While some studies have highlighted accuracy issues with OpenAlex's metadata—such as problems with institutions (Zhang et al., 2024), author names (Mongeon et al., 2023), and languages (Céspedes et al., 2024)—few have examined the quality of its national coverage. Our analysis focuses on the distinctive trend that appears in the data concerning publications from China, revealing significant insights into the limitations and challenges of using OpenAlex for country-level bibliometric analyses. The fluctuation in the number of publications indexed for China is a trend that is absent in other bibliographic databases like WoS and Scopus. This irregularity, which primarily affects journal articles, suggests that OpenAlex's coverage may be incomplete or inconsistent, especially for China's publications. This observation is particularly concerning for studies that rely on longitudinal analysis, as such data inconsistencies could lead to inaccurate interpretations of national research productivity, impact, and collaboration.

A major contributor to this fluctuation is OpenAlex's unstable indexing of publications from CNKI, a critical source for Chinese-language research outputs. Our results show that a significant proportion of China's journal articles sourced from CNKI were indexed by OpenAlex in earlier years but gradually declined to nearly zero by 2016. According to our survey, this may be attributed to increased access restrictions imposed by CNKI on overseas users, which may have hindered automated crawling or data harvesting by OpenAlex or its predecessor, MAG. While we do not have access to the proprietary ingestion pipelines of OpenAlex or MAG, nor to CNKI's backend systems, our observations—supported by publication trend analyses and metadata inspection—offer circumstantial evidence that ingestion instability is a primary factor underlying the abnormal trend in China's publication counts. This loss of coverage undermines OpenAlex's utility for accurately capturing the scholarly output of one of the world's most prolific research countries. Additionally, the fact that many CNKI papers were misidentified in terms of language and affiliation further exacerbates the reliability of OpenAlex for China's research outputs, hinting a serious intersectionality of language and country biases in major bibliographic databases.

These findings have broader implications for future studies in science of science and innovation, public policy, and quantitative science studies, especially those focusing on non-Western countries or non-English outputs. They are also relevant for researchers relying on bibliometric data, regular data users, and database developers working with open bibliographic resources. Bibliographic databases are pivotal for evaluating research productivity and informing science policies, but the biases inherent in these databases—whether due to coverage gaps or technical inaccuracies—can affect the results. For instance, OpenAlex's underrepresentation of Chinese works from CNKI could lead to an undervaluation of China's scientific contributions in global comparisons, influencing both the perception and the policymaking processes concerning global research dynamics. For bibliometric researchers, our results highlight the need for caution when using OpenAlex, particularly when assessing research outputs from countries with national databases like CNKI that may not be consistently or accurately indexed. CNKI-sourced publications could be filtered by identifying works with "primary_location"-"landing_page_url" field containing ".cnki.". It could be reasonable for data users to exclude such publications with carefully validated filtered strategies, and rely on other established databases. Moreover, OpenAlex's openness and flexibility, while advantageous in many respects, must be coupled with a clear understanding of its limitations regarding national coverage and metadata accuracy. For regular data users, querying OpenAlex by author or institution may not return a complete set of relevant works. As our findings reveal, OpenAlex exhibits metadata quality issues related to authorship. We recommend that users seeking non-English outputs cross-reference OpenAlex data with national bibliographic databases as a complementary source. For database developers—most notably the OpenAlex team—our findings point to the necessity of implementing stricter standards for source ingestion and metadata quality control, including ensuring comprehensive and sustained coverage of major national databases, where possible; implementing robust quality assurance protocols to improve the accuracy of key metadata fields, especially language and authorship.

The implications extend beyond China. Our additional analysis of South Korea, France, and Indonesia demonstrates that language misidentification is a widespread issue in OpenAlex, particularly for non-English works. This misidentification is especially pronounced in Indonesian publications, where the high error rates in language labeling could mislead analyses that assume English-language predominance in scientific outputs (Céspedes et al., 2024). Therefore, OpenAlex's data should be cross-validated with other sources, especially for non-English-speaking countries, to avoid misinterpretation of national research output and trends.

In conclusion, while OpenAlex offers valuable advantages, particularly in terms of openness and inclusivity, its limitations must be critically evaluated in bibliometric studies. Future research should consider complementing OpenAlex data with other databases or conducting manual validation of key metadata fields, especially for non-English works, to ensure comprehensive and accurate assessments. Understanding these nuances is crucial for advancing the field of science of science and innovation, particularly as researchers seek to expand the global representation and understanding of scientific productivity.

This study has some methodological limitations. First, our empirical exploration relies on the identification of institutions and country affiliation information of institutions provided by OpenAlex. However, a substantial number of works in OpenAlex lack country information (Alperin et al., 2024); meanwhile, missing institutions in OpenAlex may lead to underrepresentation of certain countries' publications (Zhang et al., 2024). Second, our analysis is limited to publications labelled as "English" language, which may underestimate the national coverage of works in other languages. Moreover, we notice that there is about 13.2% of works labelled as "article" having no source type and about 79% having no publication version for China's publications. Thus, the current paper may bias in terms of source type and versions of publications. These issues call for future in-depth analyses on the collection, analysis, and application of OpenAlex data.

**Acknowledgments**
Yi Bu acknowledges the financial supports from the National Natural Science Foundation of China (#72474009, #72104007, and #72174016) and from the 2024 Cultural Research Project of Ningbo (#WH24-2-4). Mengxue Zheng acknowledges the financial support from the China Scholarship Council (#202206270090).

# Appendix A

**Table A1**. Chi-square test results comparing CNKI and non-CNKI publications labeled as English by OpenAlex (with at least one Chinese-affiliated author), based on manual verification of full-text language

| Sample size | Verified Language | number | | χ2 | P |
| --- | --- | --- | --- | --- | --- |
| | | From CNKI | Not from CNKI | | |
| 1,634 | English | 36 | 855 | 1327.15 | .000 |
| | Chinese | 700 | 43 | | |

Note: CNKI publications are randomly sampled (100 per year) from 2000, 2003, 2006, 2009, 2011, 2013, and 2016; all 36 available CNKI records from 2020 are included. Non-CNKI publications are randomly sampled (100 per year) from 2000, 2003, 2006, 2009, 2011, 2013, 2016, 2020, and 2023.